\documentclass[twocolumn,prb,superscriptaddress,showpacs,epsf]{revtex4}
\usepackage{graphicx}
\usepackage{epsf}
\usepackage{amsmath}
\usepackage[dvips]{color}
\usepackage{pifont}
\input{colordvi.sty}

\begin{document}
\definecolor{blue}{rgb}{0.15,0.1,0.7}
\definecolor{red}{rgb}{0.7,0.1,0.15}
\definecolor{green}{rgb}{0.15,0.7,0.15}
\newcommand{\blue}[1]{{\textcolor{blue}{ \it\small #1}}}
\newcommand{\add}[1]{{\textcolor{red}{#1}}}
\title{Backaction of a charge detector on a double quantum dot}
\date{\today}
\author{Shi-Hua Ouyang}
\affiliation{Department of Applied Physics, Hong Kong Polytechnic
University, Hung Hom, Hong Kong, China} \affiliation{Department of
Physics and Surface Physics Laboratory (National Key Laboratory),
Fudan University, Shanghai 200433, China}
\author{Chi-Hang Lam}
\affiliation{Department of Applied Physics, Hong Kong Polytechnic
University, Hung Hom, Hong Kong, China}
\author{J. Q. You}
\affiliation{Department of Physics and Surface Physics Laboratory
(National Key Laboratory), Fudan University, Shanghai 200433, China}

\begin{abstract}
We develop a master equation approach to study the backaction of
quantum point contact (QPC) on a double quantum dot (DQD) at zero
bias voltage. We reveal why electrons can pass through the zero-bias
DQD only when the bias voltage across the QPC exceeds a threshold
value determined by the eigenstate energy difference of the DQD.
This derived excitation condition agrees well with experiments on
QPC-induced inelastic electron tunneling through a DQD [S.
Gustavsson {\it et al.}, Phys. Rev. Lett. \textbf{99},
206804(2007)]. Moreover, we propose a scheme to generate a pure spin
current by the QPC in the absence of a charge current.
\end{abstract}
 \maketitle

\section{Introduction}

Recent technological advances have made it possible to confine,
manipulate, and measure a small number of electrons or just one
electron in a single or double quantum dot
(DQD).\cite{ElzermanNature04,PettaPRL04,DiCarloPRL04,HayashiPRL03}
In most experiments, the electron occupancy in a quantum dot is
usually measured by a local quantum point contact (QPC) charge
detector.\cite{FieldPRL93,ElzermanPRB03} In such a system, the
backaction of the charge detector to the DQD is of particular
interest.

\begin{figure}
\includegraphics[width=2.50in,bbllx=258,bblly=138,bburx=578,bbury=458]
{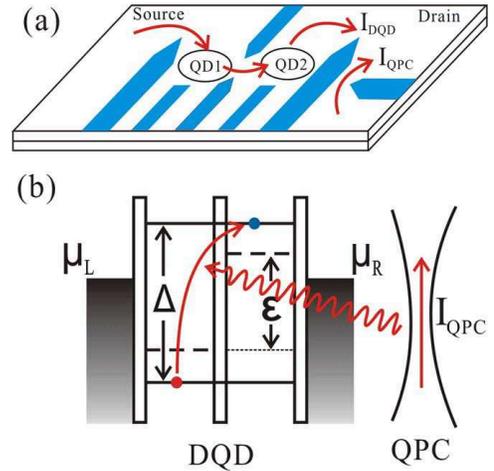} \caption{(Color online)~(a) Schematic diagram of a DQD
connected to two electrodes via tunneling barriers. A QPC used for
measuring the electron states yields backaction on the DQD. (b)
QPC-driven interdot electronic transition between two DQD
eigenstates with an energy difference $\Delta$. The energy detunning
$\varepsilon$ between the two single-dot levels (dashed lines) can
be varied by tuning the gate voltages.}\label{fig1}
\end{figure}

Several previous theoretical works (see e.g.,
Refs.~\onlinecite{GoanPRB01,GurvitzPRB97,Ouyang06}) involving this
coupled DQD-QPC system mainly focus on the QPC-induced decoherence
of the electronic states in the DQD. Recently, the impacts of the
backaction on the electron transport through a zero-bias DQD were
experimentally investigated.\cite{Gustavsson07,Gasser09} It is
suggested that the QPC emits photons which can be absorbed by a
nearby zero-bias DQD. The photon absorption process at the same time
induces the interdot electronic transitions inside the DQD and then
changes the DQD occupancy, which can be measured by the QPC. These
works show how strong the backaction of a detector on a qubit is and
provide an efficient solid state implementation for the detection of
a single photon. More importantly, the experiments show that these
interdot transitions can only be driven when the energy $|eV_d|$
(with $e$ the charge unit and $V_d$ the bias voltage across the QPC)
emitted by the QPC exceeds the eigenenergy difference $\Delta$ of
the DQD, rather than the energy difference $\varepsilon$ of the
local orbital levels in the two dots [see Fig.~\ref{fig1}(b)]. This
means that, if $|eV_d|\leq\Delta$, the interdot transitions cannot
be driven. Previously suggested mechanisms based on current
fluctuations through the QPC for interpreting the inelastic
transition in Ref.~\onlinecite{Gustavsson07} involve a perturbative
approximation \cite{Gustavsson07,AguadoPRL,Onac06} which is valid
for a weak interdot coupling. An alternative
mechanism\cite{Gasser09,Khrapai06} considering the QPC as an
effective bosonic bath of the DQD was also proposed to describe the
underlying physics. However, how this effective bosonic bath is
related to the QPC is not explicitly demonstrated.

In this paper, we theoretically analyze the backaction of the QPC on
the DQD. Starting from a microscopic description of the whole
system, we derive a master equation (ME) based on the
eigenstate-basis of the DQD to describe
the quantum dynamics of the DQD. 
Similar eigenstate-basis ME was used by Stace and
Barret\cite{StacePRL04} to study the QPC-induced decoherence
properties of the DQD. We show that this ME approach provides a
satisfactory theoretical understanding of the backaction of the QPC
on the DQD. In particular, experimental observations of the
inelastic electron tunneling through a zero-bias DQD driven by a
nearby QPC\cite{Gustavsson07,Gasser09} can be well explained.
Moreover, we propose an approach for generating a pure spin current
through a DQD. Interestingly, this spin current is driven by the
nearby QPC and can occur without a charge current.

This paper is organized as follows. In Sec.~II, we model the coupled
DQD-QPC system and derive a master equation to describe the quantum
dynamics of the DQD in the presence of the charge detector. As shown
in Sec.~III, this master equation naturally yields the main
condition under which the electron in the DQD can be excited by the
QPC. In Sec.~IV, we show that the current through the DQD can be
induced by this QPC even when the DQD is at zero-bias voltage.
Specifically, this QPC-induced current is proportional to the
capacitive coupling strength between the DQD and the QPC. In Sec.~V,
based on the same excitation mechanism by a nearby QPC, we propose a
scheme to generate a pure spin current through a zero-bias DQD.
Finally, we conclude in Sec.~VI.

\section{Characterization of a DQD coupled to a QPC}

For a DQD, both intra- and inter-dot Coulomb repulsions play an
important role in the Coulomb-blockade effect (see, e.g.,
Ref.~\onlinecite{YouPRB}). Here we consider the regime with strong
intra- and inter-dot Coulomb interactions so that only one electron
is allowed in the DQD (see Fig.~\ref{fig1}). The states of the DQD
are denoted by the occupation states $|0\rangle$, $|1\rangle$, and
$|2\rangle$, representing respectively an empty DQD, one electron in
the left dot, and one electron in the right dot. The total
Hamiltonian of the system is given by
\begin{equation}
H_{\rm tot}=H_{\rm DQD} + H_{\rm QPC} + H_{\rm det} + H_{\rm
leads} + H_{\rm T},
\end{equation}
where (we set $\hbar=1$)
\begin{subequations}
\begin{align}
&H_{\rm DQD}\! =\! \frac{\varepsilon}{2}\sigma_z+\Omega\sigma_x,\label{HDQD}\\
&H_{\rm QPC}\! =\! \sum_{kq}\omega_{Sk}c_{Sk}^{\dagger}c_{Sk}+\omega_{Dq}c_{Dq}^{\dagger}c_{Dq},\label{HQPC}\\
&H_{\rm det}\! =\! \sum_{kq}(T-\chi_1a_1^\dagger a_1-\chi_2a_2^\dagger a_2)
(c_{Sk}^\dagger c_{Dq}+{\rm H.c.}),\label{Hdet}\\
&H_{\rm leads}\! =\!  \sum_{s}\omega_{ls}c_{ls}^{\dagger}c_{ls}+\omega_{rs}c_{rs}^{\dagger}c_{rs},\\
&H_{\rm T}  \! =\!  \sum_{s}(\Omega_{ls}c_{ls}^{\dagger}a_1
+\Omega_{rs}\Upsilon_r^{\dagger}c_{rs}^{\dagger}a_2+{\rm H.c.}).\label{HT}
\end{align}
\end{subequations}
Here $H_{\rm DQD}$, $H_{\rm QPC}$, and $H_{\rm leads}$ are the free
Hamiltonians of the DQD, the QPC, and the electrodes connected to
the DQD respectively. In Eq.~(\ref{HDQD}), $\varepsilon$ is the
energy detuning between the two dots and $\Omega$ is the interdot
coupling. $c_{S k}$ ($c_{Dk}$) is the annihilation operator for
electrons in the source (drain) of the QPC with momentum $k$, while
$c_{\alpha s}$ is the annihilation operator for electrons in the
$\alpha$th ($\alpha=l,r$) electrode. Moreover, $\sigma_z=a_2^\dagger
a_2-a_1^\dagger a_1$ and $\sigma_x=a_2^\dagger a_1+a_1^\dagger a_2$
are Pauli matrices, with $a_1$ ($a_2$) the annihilation operator for
electrons staying at the left (right) dot. $H_{\rm det}$ describes
the electrostatic DQD-QPC coupling, in which $T$ is the tunneling
amplitude of an isolated QPC and $\chi_{1}$ ($\chi_2$) gives the
variation of the tunneling amplitude when the extra electron stays
at the left (right) dot. Usually one has $\chi_1<\chi_2$ since the
QPC is located more closely to the right dot. Furthermore, $H_{\rm
T}$ gives the tunneling couplings of the DQD to the two electrodes
and it depends on the tunneling coupling strengths $\Omega_{lk}$ and
$\Omega_{rk}$. In Eq.~(\ref{HT}), we have also introdued the
operators $\Upsilon_r$, $\Upsilon_r^\dagger$ to count the number of
electrons that have tunneled into the right lead.\cite{ME-E}

First, we diagonize the Hamiltonian of the DQD as
\begin{eqnarray}
H_{\rm DQD}=\frac{\Delta}{2}(|e\rangle\langle e|-|g\rangle\langle g|)=\frac{\Delta}{2}\sigma_z^{(e)},
\end{eqnarray}
where $\Delta=\sqrt{\varepsilon^2+4\Omega^2}$ is the energy
splitting of the two eigenstates of the DQD given by $|g\rangle =
\alpha|1\rangle-\beta|2\rangle$, and $|e\rangle
=\beta|1\rangle+\alpha|2\rangle$, with
$\alpha\equiv\cos{\theta}/{2}$, $\beta\equiv\sin{\theta}/{2}$, and
$\tan\theta=2\Omega/\varepsilon$.
This eigenstate basis will be adopted in all the following
calculations. To describe the dynamics of the system, we derive the
ME starting from the von Neumann equation
\begin{equation}
\dot{\rho}_R(t)=-i[H_{\rm tot},\,\rho_R],
\end{equation}
with $\rho_R$ the density matrix of the whole system. In the
interaction picture defined by the free Hamiltonian
\begin{equation}
H_0=H_{\rm DQD} + H_{\rm QPC} + H_{\rm leads},
\end{equation}
the interaction
Hamiltonian $H_{\rm I}=H_{\rm det}+H_{\rm T}$ is given by 
\begin{eqnarray}
H_{\rm I}(t) = A(t)Y(t)+H_{\rm T}(t),\label{Hint}
\end{eqnarray}
where
\begin{subequations}
\label{Hint-2}
\begin{align}
A(t)\!& = \! \sum_{n=1}^3P_ne^{-i\omega_nt},\label{A(t)}\\
Y(t) \!& = \! \sum_{kq}[c_{Sk}^\dagger c_{Dq}e^{i(\omega_{Sk}-\omega_{Dq})t}+{\rm H.c.}],
\label{Hint-t}\\
H_{\rm T}(t) \!& = \! \sum_s\big[c_{ls}^\dagger(\alpha a_ge^{-i\Delta t/2}+\beta a_ee^{i\Delta t/2})e^{i\omega_{ls}t}
+\Upsilon_r^\dagger c_{rs}^\dagger
\nonumber\\
&\times(-\beta a_ge^{-i\Delta t/2t}+\alpha a_ee^{i\Delta t/2})e^{i\omega_{rs}t}+{\rm H.c.}\big].
\end{align}
\end{subequations}
In Eq.~(\ref{A(t)}), we have defined
\begin{eqnarray}
&&P_1=-\alpha\beta|e\rangle\langle g|,~~P_2=-\alpha\beta|g\rangle\langle e|,\nonumber\\
&&P_3=T-(\chi_1|e\rangle\langle e|+\chi_2|g\rangle\langle
g|)+\alpha^2\chi_d\sigma_z^{(e)},
\end{eqnarray}
with $\chi_d=\chi_1-\chi_2$. Also, we have chosen
$\omega_1=-\omega_2=-\Delta$, and $\omega_3=0$. Integrating the von
Neumann equation within the Born-Markov approximation and tracing
over the degrees of freedom of the QPC and the two electrodes, one
obtains\cite{Blum}
\begin{eqnarray}
\dot{\rho}(t)\!&\!=\!&\!{\rm Tr}_{R}\bigg(\!\!-i[H_{\rm I}(t), \rho_R(0)]
\nonumber\\&&
-\!\int\limits_0^tdt'[H_{\rm I}(t),\,[H_{\rm I}(t'),\,\rho_R(t)]]\bigg).\label{ME-integral}
\end{eqnarray}
Following the experimental conditions, we consider a zero bias
across the DQD. We also set $\mu_L=\mu_R=0$ [see Fig.~\ref{fig1}(b)]
although other values of $\mu_L$ and $\mu_R$ satisfying
$\Delta/2\!>\!\mu_L\!>\!-\Delta/2$ give identical results .
Substituting $H_{\rm I}(t)$ shown in Eqs.~(\ref{Hint}) and
(\ref{Hint-2}) into Eq.~(\ref{ME-integral}), neglecting the
fast-oscillating terms (which are proportional to $e^{\pm i\Delta
t}$), and converting the obtained equation into the Schr\"{o}dinger
picture, we obtain the ME:\cite{ME-E}
\begin{eqnarray}
&&\dot{\rho}(t)\!=\!-i[H_{\rm DQD},\,\rho(t)]+\mathcal{L}_d\rho(t)+\mathcal{L}_T\rho(t)
+\gamma\mathcal{D}[a_g^\dagger a_e]\rho,\nonumber\\
&&~~\label{ME}
\end{eqnarray}
with
\begin{eqnarray}
\mathcal{L}_d\rho(t)\!&\!=\!&\!2\pi g_Sg_D\sum_{i=1}^3\big\{\mathcal{D}[P_i]\rho(t)\Theta(eV_d+\omega_i)
\nonumber\\&&+\mathcal{D}[P_i^\dagger]\rho(t)\Theta(-eV_d-\omega_i)\big\},
\nonumber\\
\mathcal{L}_T\rho(t)\!&\!=\!&\!\Gamma_L\alpha^2\mathcal{D}[a_g^\dagger]\rho+\Gamma_L\beta^2\mathcal{D}[a_e]\rho
\nonumber\\
&&+\Gamma_R\beta^2\mathcal{D}[a_g^\dagger\Upsilon_r]\rho+\Gamma_R\alpha^2\mathcal{D}[a_e\Upsilon_r^\dagger]\rho.
\label{ME-L}
\end{eqnarray}
The notation $\mathcal{D}$ acting on any operator $A$ is defined by
\begin{equation}
\mathcal{D}[A]\rho=A\rho A^\dagger-\frac{1}{2}(A^\dagger A\rho+\rho
A^\dagger A).
\end{equation}
Also, $\Gamma_{L(R)}=2\pi g_{L(R)}\Omega_{lk(rk)}^2$ is the rate for
electron tunneling through the left (right) barrier and
$\Theta(x)=(x+|x|)/2$. $g_{i}$ ($i=S$,$D$,$L$, or $R$) denotes the
density of states at the QPC source, the QPC drain, the left DQD
electrode, or the right DQD electrode, which is assumed to be
constant over the relevant energy range. Furthermore, to allow for
the couplings of the DQD to other degrees of freedom, such as
hyperfine interactions and electron-phonon couplings, we
phenomenologically include an additional relaxation term [the fourth
term in Eq.~(\ref{ME})] describing transitions from the excited
state $|e\rangle$ to the ground state $|g\rangle$.

\section{QPC-induced excitation condition}

Below we use the ME to derive the electron tunneling rates into and
out of the DQD at zero bias voltage. From Eqs.~(\ref{ME}) and
(\ref{ME-L}) and the relations
\begin{eqnarray}
&&\langle n|\Upsilon_r^\dagger\Upsilon_r\rho|n\rangle=\rho^{(n)},~\langle n|\Upsilon_r\Upsilon_r^\dagger\rho|n\rangle=\rho^{(n)},
\nonumber\\&&
\langle n|\Upsilon_r^\dagger\rho\Upsilon_r|n\rangle=\rho^{(n-1)},~\langle n|\Upsilon_r\rho\Upsilon_r^\dagger|n\rangle=\rho^{(n+1)},
\end{eqnarray}
we obtain the $n$-resolved equation of motion for each reduced
density matrix element:
\begin{eqnarray}
\dot{\rho}_{00}^{(n)} \!& = \!& -\Gamma_1\rho_{00}^{(n)}
+\Gamma_L\beta^2\rho_{ee}^{(n)}+\Gamma_R\alpha^2\rho_{ee}^{(n-1)},\label{EOM-a}\nonumber\\
\dot{\rho}_{gg}^{(n)} \!& = \!& \Gamma_L\alpha^2\rho_{00}^{(n)}+\Gamma_R\beta^2\rho_{00}^{(n+1)}
-\Gamma_a\rho_{gg}^{(n)}+(\Gamma_e+\gamma)\rho_{ee}^{(n)},\label{EOM-b}\nonumber\\
\dot{\rho}_{ee}^{(n)} \!& = \!& -\Gamma_2\rho_{ee}^{(n)}
+\Gamma_a\rho_{gg}^{(n)}-(\Gamma_e+\gamma)\rho_{ee}^{(n)},\label{EOM}
\end{eqnarray}
where $n$ is the number of electrons that have tunneled through the
DQD via the right tunneling barrier at time $t$, so that
$\rho_{ij}=\sum_n\rho_{ij}^{(n)}$ $(i,j\!=\!0,g,e)$. In
Eq.~(\ref{EOM}), we have defined the QPC-induced excitation and
relaxation rates:
\begin{equation}
\Gamma_{a,e}=2\pi
g_Sg_D\chi_d^2(\alpha\beta)^2[\Theta(eV_d\mp\Delta)+\Theta(-eV_d\mp\Delta)].
\end{equation}
as well as the tunneling rates
$\Gamma_1=\Gamma_L\alpha^2+\Gamma_R\beta^2$, and
$\Gamma_2=\Gamma_L\beta^2+\Gamma_R\alpha^2$.
\begin{figure}[tbp]
\includegraphics[width=2.50in,bbllx=11,bblly=48,bburx=242,bbury=378]
{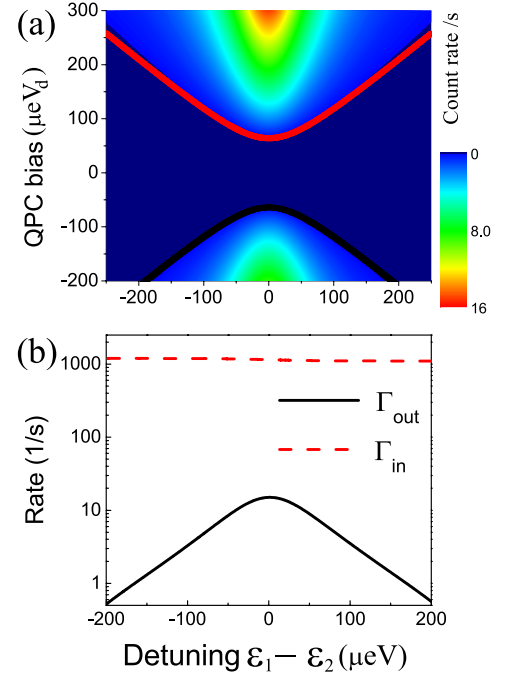} \caption{(Color online) (a) Tunneling rate $\Gamma_{\rm
out}$ as a function of both the energy detuning
$\varepsilon_1-\varepsilon_2$ and the QPC bias voltage $V_d$. The
hyperbolic curves mark the two eigenenergy levels of the DQD. (b)
Tunneling rates $\Gamma_{\rm in}$ and $\Gamma_{\rm out}$ versus the
energy detuning. We use the typical experimental
parameters in Ref.~\onlinecite{Gustavsson07}: $V_d=-300~\mu$eV, $\Omega=32~\mu$eV,
$\Gamma_L=1.2$~kHz, $\Gamma_R=1.1$~kHz, $T=0.5$, $\chi_1=0.028~T$,
$\chi_2=0.05~T$, and $1/\gamma=2$~ns.}\label{fig2}
\end{figure}

The environment-induced relaxation rate\cite{PettaPRL04} $\gamma$
($\!\sim\!1~$ns) of a DQD is typically much larger than the
available measurement bandwith of the QPC, so that transitions
between the ground state $|g\rangle$ and the excited state
$|e\rangle$ cannot be directly registered by the detector. However,
using time-resolved charge-detection techniques, one can measure
whether the DQD is occupied by an extra electron or not, so that the
measured time traces show two levels.\cite{Gustavsson07} From these
traces, one can extract the rates of electrons tunneling into and
out of the DQD, i.e., $\Gamma_{\rm in}=1/\langle\tau_{in}\rangle$
and $\Gamma_{\rm out}=1/ \langle\tau_{out}\rangle$ with $\tau_{in}$
and $\tau_{out}$ being the waiting times for the corresponding
tunneling events. At steady state with
$\dot{\rho}_{ij}(t\rightarrow\infty)=0$, the number of electrons
coming from the left lead to the DQD should be equal to the number
of electrons going out of the DQD to the right lead. Thus, one has
the equilibrium relation:
\begin{eqnarray}
\Gamma_{\rm in}\rho_{00}=\Gamma_{\rm out}(\rho_{gg}+\rho_{ee}).
\end{eqnarray}
For an initially empty DQD, $|0\rangle$, the first term of the first
equation in Eq.~(\ref{EOM-a}) implies that $\Gamma_{\rm
in}=\Gamma_1$. Using also the steady state solution of
Eq.~(\ref{EOM}), straightforward algebra gives
\begin{equation}
\Gamma_{\rm
out}\!=\!\frac{\Gamma_a\Gamma_2}{\Gamma_2+\Gamma_e+\gamma}.\label{Fout-1}
\end{equation}
As expected, the rate for electrons tunneling out of the DQD is
proportional to the QPC-induced excitation rate $\Gamma_a$. Using
$\Gamma_a$ obtained above, it follows from Eq.~(\ref{Fout-1}) that
\begin{eqnarray}
\Gamma_{\rm out}\!&\!=\!&\!\frac{2\pi
g_Sg_D\chi_d^2(\alpha\beta)^2\Gamma_2[\Theta(eV_d-\Delta)+\Theta(-eV_d-\Delta)]}{\Gamma_2+\Gamma_e+\gamma}.
\nonumber\\&&
\label{Fout}
\end{eqnarray}
%
We emphasize that this excitation process occurs (i.e., $\Gamma_{\rm
out}>0$) only when the energy emitted by the QPC is larger than the
eigenenergy difference of the DQD, i.e.,
\begin{equation}
|eV_d|>\Delta.
\end{equation}
Otherwise, if $|eV_d|\leq\Delta$, $\Gamma_{\rm out}$ is always zero
and no excitation process happens. This agrees well with the
experiment in Ref.~\onlinecite{Gustavsson07}. More importantly, we
note from Eq.~(\ref{Fout}) that
\begin{equation}
\Gamma_{\rm out}\propto\chi_d^2=(\chi_1-\chi_2)^2,
\end{equation}
which clearly implies that the QPC-induced excitation results from
the electrostatic coupling between the DQD and the QPC. Figure
\ref{fig2}(a) plots the rate $\Gamma_{\rm out}$ as a function of
both the QPC-bias voltage $V_d$ and the energy detunning
$\varepsilon_1-\varepsilon_2$ of the DQD. Figure \ref{fig2}(b) gives
the calculated rates for electrons tunneling into and out of the DQD
as a function of the energy detunning. For almost symmetric
tunneling couplings, i.e., $\Gamma_L\simeq\Gamma_R$, the rate
$\Gamma_{\rm in}$ is almost constant and $\Gamma_{\rm out}$ depends
strongly on the energy detunning.

\section{QPC-driven charge current through the DQD}

Without a nearby QPC, an electron can enter the empty zero-bias DQD,
but should be trapped by the DQD. This leads to zero current through
the DQD. However, the QPC  can induce an excitation from the ground
state $|g\rangle$ to the excited state
$|e\rangle$~[Fig.~\ref{fig1}(b)], from which the electron leaves the
DQD. When this cycle repeats, a finite current flows through the
DQD. Below we study this QPC-induced charge current when the
couplings between the DQD and the two electrodes are strongly
asymmetric, i.e., $\Gamma_L\gg\Gamma_R$, as investigated by Gasser
{\it et al.}.\cite{Gasser09} The charge current $I_c(t)$ through the
DQD at time $t$ is
\begin{equation}
I_c(t) = e\frac{dN(t)}{dt} =e\sum_{n,i}n\dot{\rho}_{ii}^{(n)},
\end{equation}
where $N(t)$ is the number of electrons that have tunneled into the
right lead. From Eq.~(\ref{EOM}), we get
\begin{eqnarray}
I_c(t)=e\Gamma_R(\alpha^2\rho_{ee}-\beta^2\rho_{00}).
\end{eqnarray}
At steady state, the current becomes
\begin{eqnarray}
I_c=\frac{e\Gamma_L\Gamma_R\Gamma_a(\alpha^2-\beta^2)}
{\Gamma_a(\Gamma_L+\Gamma_R)+\Gamma_1(\Gamma_e+\gamma+\Gamma_2)}.\label{cc}
\end{eqnarray}
In Gasser {\it et al.}'s experiment, the DQD is coupled to two QPCs,
i.e., one QPC is adjacent to the left dot and the other to the right
dot. This can be characterized by replacing $H_{\rm QPC}$ and
$H_{\rm det}$ in Eqs.~(\ref{HQPC}) and (\ref{Hdet}) by
\begin{eqnarray}
H_{\rm QPC}\!&\!=\!&\!\sum_{i}H_{\rm QPC}^i\!=\!
\sum_{kqi}\omega_{Ski}c_{Ski}^{\dagger}c_{Ski}+\omega_{Dqi}c_{Dqi}^{\dagger}c_{Dqi},
\nonumber\\
H_{\rm det} \!&\!=\!&\!\sum_{kqi}(T_i-\chi_{1i}a_1^\dagger
a_1-\chi_{2i}a_2^\dagger a_2) (c_{Ski}^\dagger c_{Dqi}+{\rm H.c.}).
\nonumber\\&&
\end{eqnarray}
Here $H_{\rm QPC}^i$ is the Hamiltonian of the $i$th ($i=L,R$) QPC
and $\chi_{1i}$ ($\chi_{2i}$) gives the variation of the tunneling
amplitude of the $i$th QPC when the extra electron stays in the left
(right) dot. From Eq.~(\ref{cc}), one notes that the charge current
is determined by $\chi_d^2$, which characterizes the coupling
strength between the DQD and the QPC.
\begin{figure}
\includegraphics[width=3.20in,bbllx=18,bblly=48,bburx=338,bbury=258]{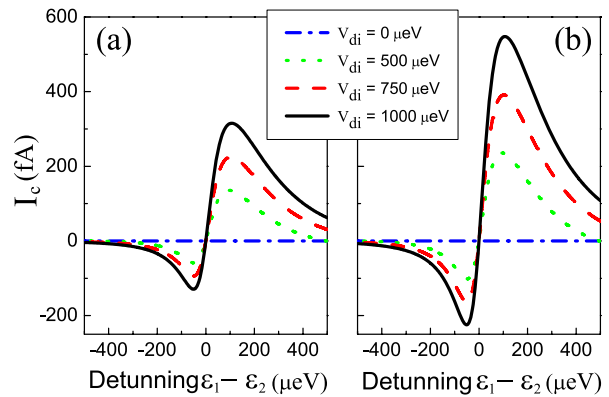}
\caption{(Color online)~QPC-driven
static charge current $I_c$ through the DQD as a function of the
energy detuning for different QPC voltages. (a)
$V_{d1}\in[0,1000]~\mu$eV, $V_{d2}=0$, and
$\chi_{d1}=\chi_{1L}-\chi_{2L}=0.022~T_1$; (b)
$V_{d2}\in[0,1000]~\mu$eV, $V_{d1}=0$, and
$\chi_{d2}=\chi_{1R}-\chi_{2R}=-0.029~T_2$. Other parameters are
$\Gamma_L=0.5~$GHz, $\Gamma_R=7.8~$MHz, $\Omega=35~\mu$eV,
$T_1=T_2=0.5$,
and $1/\gamma=5~$ns. }\label{fig3}
\end{figure}

In Fig.~\ref{fig3}, we plot the charge current versus energy
detunning for various QPC voltages. Here we choose the current
direction from the left to the right to be positive. Comparing
Fig.~\ref{fig3}(a) with Fig.~\ref{fig3}(b), one sees that the
current is more pronounced for a larger value of $\chi_d^2$. This is
because the QPC-induced excitation becomes stronger when $\chi_d^2$
is larger. Furthermore, the asymmetry of the current with respect to
the energy detuning $\varepsilon_1-\varepsilon_2$ can be explained
as follows. When $\varepsilon_1>\varepsilon_2$, one has
$\alpha<\beta$ and electrons tunnel from the right electrode to the
left one via the DQD. In contrast, electrons tunnel in the opposite
direction when $\varepsilon_1<\varepsilon_2$. In the latter case,
however, a small tunneling rate $\Gamma_R$ will partially block the
current, which leads to the asymmetry of the current. These features
were clearly observed in the experiment.\cite{Gasser09}

\section{Novel spin-current generator}

Finally, we propose a new scheme to generate a pure spin current
through the DQD by applying two different static magnetic fields in
the quantum dots. Unequal magnetic fields in the two dots can be
realized by placing a Co micromagnet near one dot of the
DQD.\cite{NeilPRL08} When the eigenstate energy difference $\Delta$
becomes spin dependent, the QPC-driven excitation rate depends on
the spin states. A spin-polarized current through the DQD can be
generated. Here we consider the case with a static magnetic field
$B_z$ applied only at the left dot. Generalizing the ME in
Eq.~(\ref{EOM}) to take into account the spin degrees of freedom,
the current due to electrons with spin $\sigma$
($=\uparrow,\downarrow$) is given by
\begin{equation}
I_{\sigma}=\frac{e\Gamma_L\Gamma_R\Gamma_a^{\sigma}
\Gamma_a^{\bar{\sigma}}\Gamma_2^{\bar{\sigma}}(\alpha_\sigma^2-\beta_\sigma^2)}
{M_\sigma},\label{sc}
\end{equation}
where
\begin{eqnarray}
M_\sigma\!&=&\!\Gamma_a^\sigma\Gamma_a^{\bar{\sigma}}
(\Gamma_2^\sigma\Gamma_2^{\bar{\sigma}}+\Gamma_1^\sigma
\Gamma_2^{\bar{\sigma}}+\Gamma_1^{\bar{\sigma}}\Gamma_2^\sigma)
\nonumber\\
&&+\sum_{s=\sigma,\bar{\sigma}}\Gamma_a^s\Gamma_1^{\bar{s}}\Gamma_2^s
(\Gamma_2^{\bar{s}}+\Gamma_2^{\bar{s}}+\gamma).
\end{eqnarray}
All the spin-related parameters in Eq.~(\ref{sc}), e.g.,
$\tan\theta_\sigma=2\Omega/\varepsilon_\sigma$ and
$\Gamma_1^\sigma=\Gamma_L\alpha_\sigma^2+\Gamma_R\beta_\sigma^2$,
are calculated by simply replacing $\varepsilon$ in the original
expression by
\begin{eqnarray}
\varepsilon_{\uparrow,\downarrow}=\varepsilon\pm{1\over2}\Delta_z,
\end{eqnarray}
where $\Delta_z=g\mu_BB_z$ is the Zeeman splitting with $g$ being
the $g$ factor and $\mu_B$ the magneton.
\begin{figure}
\includegraphics[width=2.80in,bbllx=16,bblly=53,bburx=218,bbury=228]{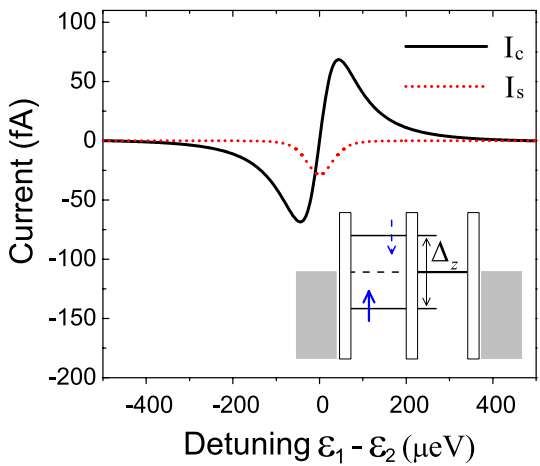}
\caption{(Color online) QPC-induced charge ($I_c=I_\uparrow+I_\downarrow$)
and spin ($I_s=I_\uparrow-I_\downarrow$) currents through a DQD when
a static magnetic field is applied at the left dot. Here
$\Gamma_L=\Gamma_R=7.5~$MHz and the Zeeman splitting is $\Delta_z=20~\mu$eV.
The other parameters are the same as those in Fig.~\ref{fig3}(a).
Inset: Schematic diagram of the energy levels in the DQD at $\epsilon_1 = \epsilon_2$.}
\label{fig4}
\end{figure}

In Fig.~\ref{fig4}, the charge current
($I_c=I_\uparrow+I_\downarrow$) and the spin current
($I_s=I_\uparrow-I_\downarrow$) are plotted versus the energy
detuning $\varepsilon_1-\varepsilon_2$ when $\Gamma_L=\Gamma_R$. At
zero detuning with $\varepsilon_1 = \varepsilon_2$, the charge
current is zero and the spin current reaches its maximum. This is
because electrons with opposite spins are transported with the same
effective rate but in opposite directions. Therefore, a pure spin
current without a charge current can be driven by a QPC in this
proposed device. A charge current is also induced at non-vanishing
detuning ($|\varepsilon|>0$), but its direction is sensitive to the
sign of $\varepsilon$. Specifically, the charge current is positive
when $\varepsilon_1>\varepsilon_2$ and negative vice versa. The
underlying mechanism can be demonstrated as follows. When the energy
detuning is deviated from the zero detuning point, e.g.,
$\varepsilon_1>\varepsilon_2$, the absolute value of the spin-up
current transporting from the left to the right decreases, while
that of the spin-down current transporting from the right to the
left increases. This results in a positive charge current tunneling
through the DQD. In contrast to other schemes, including electron
spin resonance~\cite{KoppensNature06} or photon-assisted
tunneling~\cite{OosterkampNature98} in QDs, our proposal does not
require a fast and strong oscillating magnetic field. Moreover,
unlike the partially polarized spin current through a QD driven by a
spin bias~\cite{LuPRB08} or spin-orbit interaction,\cite{SunPRL07} a
pure spin current without a charge current can be generated here and
the amplitude is tunable via the voltage across the QPC.

In quantum transport experiments, the most well-established readout
devices are charge detectors. However, these charge-sensitive
devices are insensitive to spins. Because of the unique properties
of the spin current in our proposed setup, charge detectors can be
used to indirectly reveal the existence of the spin current. As
shown in Fig.~\ref{fig4}, the spin current is symmetric about the
energy detuning $\varepsilon_1-\varepsilon_2$, but the charge
current has an asymmetric dependence. If the measured charge current
is demonstrated to be asymmetric about $\varepsilon$, as predicted
in Fig.~\ref{fig4},
it serves as a characteristic feature indicating that a pure spin
current occurs at the zero detuning point in our present proposal.

\section{Conclusion}
In our approach, the current through the zero-bias DQD is due to the
backaction of the QPC. This mechanism predicts that the rate for
electrons tunneling out of the zero-bias DQD is proportional to the
bias voltage applied across the QPC [see Eq.~(\ref{Fout})]. However,
in Ref.~\onlinecite{Gasser09}, where the DQD is assumed to be
coupled to acoustic phonons [see Eq.~(A4) in
Ref.~\onlinecite{Gasser09}], this tunneling rate has a polynomial
relation with the bias voltage across the QPC. We suggest that,
using the setup fabricated in Ref.~\onlinecite{Gasser09}, one can
measure the rate for electrons tunneling out of the DQD as a
function of the bias voltage across the QPC to distinguish between
these two mechanisms.

In conclusion, we have developed a ME approach to study the
backaction of a charge detector (QPC) on a DQD. We show that an
electron in the DQD can be excited from its ground state to the
excited state when the energy emitted by the QPC exceeds the
eigenenergy difference. This agrees well with the observations in
two recent experiments.\cite{Gustavsson07,Gasser09} Moreover, we
propose a new scheme to drive a pure spin current by a QPC in the
absence of a charge current.

\begin{acknowledgments}
This work is supported by the National Basic Research Program of
China Grant Nos. 2009CB929300 and 2006CB921205, the National Natural
Science Foundation of China Grant Nos. 10534060 and 10625416, and
the Research Grant Council of Hong Kong SAR project No. 500908.
\end{acknowledgments}


\end{document}